\documentstyle[preprint,aps]{revtex}
\tighten
\include{epsf}
\begin{document}
\preprint{Preprint Numbers:\parbox[t]{50mm}{ADP-96-27/T226\\
		 FSU-SCRI-96-71\\ hep-ph/9608292}}
\draft
\title{On Renormalized Strong-Coupling Quenched QED\\
        in Four Dimensions}
\author{
  Frederick T.\ Hawes \footnotemark[1],
  Tom Sizer\footnotemark[2]
	  and
  Anthony G.\ Williams\footnotemark[2]\footnotemark[3]
  \vspace*{2mm} }

\address{
  \footnotemark[1] Department of Physics and SCRI, 
    Florida State University, \\ 
    Tallahassee, Florida 32306-3016
  \vspace*{2mm}\\
  \footnotemark[2] Department of Physics and Mathematical Physics,
    University of Adelaide, \\ 5005, Australia
  \vspace*{2mm}\\
  \footnotemark[3] Institute for Theoretical Physics,
    University of Adelaide, 5005, Australia
}
%
\maketitle
%
\begin{abstract}
  We study renormalized quenched strong-coupling QED in four dimensions
  in arbitrary covariant gauge.  Above the critical coupling
  leading to dynamical chiral symmetry breaking, we show that
  there is no finite chiral limit.  This behaviour is found to be
  independent of the detailed choice of photon-fermion proper vertex
  in the Dyson-Schwinger equation formalism, provided that the vertex
  is consistent with the Ward-Takahashi identity and multiplicative
  renormalizability.  We show that the finite solutions previously
  reported lie in an unphysical regime of the theory with multiple
  solutions and ultraviolet oscillations in the mass functions.  This
  study supports the assertion that in four dimensions strong coupling
  QED does not have a continuum limit in the conventional sense.
\end{abstract}

\section{Introduction}
\label{sec_intro}

A useful approach to studying the mechanism of dynamical chiral
symmetry breaking (DCSB) is through the Dyson-Schwinger equation
(DSE) formalism \cite{TheReview,MiranskReview,FGMS}.
The infinite set of coupled DSE's must always be
truncated at some point, but we can still make progress by
closing off the tower of equations with a suitable {\it Ansatz\/}
consistent with all appropriate symmetries of the theory and having
the correct perturbative limit.  While not a complete first
principles treatment of a theory, this approach is nonetheless a
useful tool, since it does allow Lorentz covariance to
be maintained as well as allowing for the infrared (IR) and
ultraviolet (UV) limits to be numerically taken
in a straightforward way.  In addition, model independent results
following from symmetry priciples alone can still be obtained and
numerically verified in a rigorous way as we will see.
On the other hand, lattice gauge theory (LGT) studies \cite{Rothe}
are a first principles, approximation-free approach,
but they still present a significant computational challenge
when attempting to verify ultraviolet and infrared limits
\cite{TheReview,MiranskReview}.

The abelian nature of quantum electrodynamics (QED)
in many ways makes it a much simpler system to study
than a nonabelian theory such as quantum chromodynamics (QCD).
For this reason it has been the subject of many nonperturbative
studies, which have as their long-term goal a detailed understanding
of nonperturbative QCD.
On the other hand, strong-coupling QED$_4$ is widely anticipated
to behave unconventionally in the continuum limit
and for this reason is a theory of considerable interest
in its own right.

In previous work \cite{qed4_hw} we first introduced a
numerical renormalization procedure and applied it to QED$_4$ with
a quenched photon propagator using the Dyson-Schwinger formalism.
This is a direct application of the standard renormalization procedure
to the nonlinear self-consistent framework needed to study
dynamical chiral symmetry breaking.
This initial work, in Landau gauge, was recently generalized
to arbitrary covariant gauges \cite{qed4_hwr}.
The central result of these two works was to demonstrate that
the numerical renormalization procedure works extremely well
and allows the continuum limit ($\Lambda\to\infty$) to be taken
numerically, while giving rise to stable finite solutions
for the renormalized fermion propagator.

In this article we investigate the chiral limit
in renormalized quenched strong-coupling QED$_4$,
using a photon-fermion vertex
that satisfies the Ward-Takahashi Identity (WTI)
and makes the fermion DSE multiplicatively renormalizable.
We find that for couplings above the chiral critical coupling,
keeping the bare mass $m_0(\Lambda) \equiv 0$
as the cutoff is relaxed
results in a dynamical mass function with no oscillations and
which diverges with the cutoff.
The finite solutions described in previous articles
\cite{qed4_hw,qed4_hwr} showed damped oscillations
in the dynamical mass functions at large $p^2$,
which suggested that they were unphysical.
Further, we show here that
for a given supercritical coupling
and the same bare mass $m_0(\Lambda)$,
it is possible to have multiple solutions
corresponding to different renormalized masses $m(\mu)$.
We conclude that quenched strong-coupling QED in four dimensions
does not have a chiral limit in the conventional sense above
the chiral phase transition.

In Sec.\ \ref{sec_formalism} we briefly summarize
  the renormalized Dyson-Schwinger equation formalism,
  the numerical renormalization procedure,
and a particular fermion-photon vertex ansatz
that we use to illustrate our general arguments numerically.
In Sec.\ \ref{sec_results} we demonstrate the scaling of the DSE solutions 
with zero bare mass for supercritical coupling,
and the existence of multiple solutions with the same bare mass;
we also present a general argument to show that a vertex
which is consistent with the WTI and multiplicative renormalizability
leads to a diverging mass function above critical coupling
in the continuum limit.
We discuss these results and their relevance
to QCD and to unquenched QED$_4$,
in Sec.\ \ref{sec_conclusions}.
For further details and references and an expanded discussion
we refer the reader to Refs.\ \cite{TheReview,qed4_hw,qed4_hwr}. 

\section{Formalism}
\label{sec_formalism}


Dynamical chiral symmetry breaking (DCSB) occurs when the
fermion propagator develops a nonzero scalar
self-energy in the absence of an explicit chiral symmetry breaking
(ECSB) fermion mass.
We will refer to coupling constants strong enough
to induce DCSB as supercritical and those weaker as subcritical.
We write the fermion propagator as
\begin{equation}                        \label{fermprop_formal}
  S(p) = \frac{Z(p^2)}{\not\!p - M(p^2)}
       = \frac{1}{A(p^2) \not\!p - B(p^2)}
\end{equation}
where we refer to $A(p^2)\equiv 1/Z(p^2)$
as the finite momentum-dependent fermion renormalization
and where $M(p^2)\equiv B(p^2)/A(p^2)$ is the fermion mass function.
In the massless theory
(i.e., in the absence of an ECSB bare fermion mass $m_0(\Lambda)$)
by definition DCSB occurs when $M(p^2)\neq 0$. 

With the exception of Refs.~\cite{qed4_hw,qed4_hwr}, most studies
have neglected the issue of the subtractive renormalization
of the DSE for the fermion propagator.
Typically these studies have assumed an initially massless theory
and have renormalized at the ultraviolet cutoff of the loop integration,
taking $Z_1 = Z_2 = 1$.  Where a nonzero bare mass has been used,
it has simply been added to the scalar term in the propagator.
Although there were earlier formal discussions of renormalization
\cite{MiranskReview,CPI,CPII,CPIII,CPIV}, the subtractive
renormalization program had not previously been implemented.

We will concentrate our discussion on quenched strong-coupling QED$_4$,
where here the term ``quenched'' means
that the bare photon propagator is used in the fermion self-energy DSE,
so that $Z_3 = 1$ and there is no renormalization of the electron charge.
It should be carefully noted that this is a slightly different usage
from that found in lattice gauge theory studies,
since in DSE studies with a quenched photon propagator
virtual fermion loops may still be present
in the proper fermion-photon vertex.

The DSE for the renormalized fermion propagator,
in an arbitrary covariant gauge, is
\begin{equation} \label{fermDSE_eq}
  S^{-1}(p) = Z_2(\mu,\Lambda)[\not\!p - m_0(\Lambda)]
    - i Z_1(\mu,\Lambda) e^2 \int^{\Lambda} \frac{d^4k}{(2\pi)^4}
	  \gamma^{\mu} S(k) \Gamma^{\nu}(k,p) D_{\mu \nu}(q)\:;
\end{equation}
here $q=k-p$ is the photon momentum, $\mu$ is the renormalization
point, and $\Lambda$ is a regularizing parameter (taken here to be an
ultraviolet momentum cutoff).  We write
$m_0(\Lambda)$ for the regularization-parameter dependent bare mass.
The renormalized charge is $e$ (as opposed to the bare charge $e_0$),
and the general form for the renormalized photon propagator is
\begin{equation}
  D^{\mu\nu}(q) =
    \left\{
      \left( -g^{\mu\nu} + \frac{q^\mu q^\nu}{q^2} \right)
      \frac{1}{1+\Pi(q^2)}
      - \xi \frac{q^\mu q^\nu}{q^2}
    \right\}
    \frac{1}{q^2}\:,
\end{equation}
with $\xi$ the renormalized covariant gauge parameter
and $\xi_0 \equiv Z_3(\mu,\Lambda) \xi$ the corresponding bare one.
In the quenched approximation, we have for the coupling strength
and gauge parameter respectively
$\alpha\equiv e^2/4\pi = \alpha_0\equiv e_0^2/4\pi$
and $\xi=\xi_0$, and for the photon propagator we have
\begin{equation}
  D^{\mu\nu}(q) \to  D_0^{\mu\nu}(q) 
    =  \left\{\left( -g^{\mu\nu} + \frac{q^\mu q^\nu}{q^2} \right)
          - \xi \frac{q^\mu q^\nu}{q^2} \right\}\frac{1}{q^2}\:.
\end{equation}


The requirement of gauge invariance in QED leads to a set of identities
referred to as the Ward-Takahashi
Identities (WTI).  The WTI for the fermion-photon vertex is
\begin{equation}
  q_\mu \Gamma^\mu(k,p) = S^{-1}(k) - S^{-1}(p)\;,
\label{WTI}
\end{equation}
where $q = k - p\:$.  This is a generalization of the
original, differential Ward identity, which expresses the effect of
inserting a zero-momentum photon vertex into the fermion propagator,
\begin{equation}
  \frac{\partial S^{-1}(p)}{\partial p_\nu} = \Gamma^{\nu}(p,p)\:.
\label{WI}
\end{equation}
The Ward identity, Eq.~(\ref{WI}), follows immediately from the WTI of 
Eq.~(\ref{WTI}).
In general, for nonvanishing photon momentum $q$, only the longitudinal
component of the proper vertex is constrained, i.e., the WTI provides
no information on $\Gamma^\nu_{\rm T}(k,p)\equiv
{\cal T}^{\mu\nu}\Gamma_\nu(p,k)$ for $q\neq 0$.  [We use the notation
${\cal T}^{\mu\nu}\equiv g^{\mu\nu}-(q^\mu q^\nu/q^2)$ and
${\cal L}^{\mu\nu}\equiv (q^\mu q^\nu/q^2)$ for the transverse and
longitudinal projectors, respectively.]
In particular, the WTI guarantees the equality of the propagator
and vertex renormalization constants, $Z_2 \equiv Z_1$ (at least in any
sensible choice of subtraction scheme \cite{TheReview}.)
The WTI can be shown to be satisfied
order-by-order in perturbation theory and can also be derived
nonperturbatively.

As discussed in \cite{TheReview,Craig_1}, this can be thought
of as just one of a set of six general requirements on the vertex:
(i) the vertex must satisfy the WTI; (ii) it should contain no kinematic
singularities; (iii) it should transform under charge conjugation ($C$),
parity inversion ($P$), and time reversal ($T$) in the same way
as the bare vertex, e.g.,
	\begin{equation}
	  C^{-1} \Gamma_\mu(k,p) C = - \Gamma_\mu^{\rm T}(-p,-k)
	\end{equation}
(where the superscript {\rm T} indicates the transpose);
(iv) it should reduce to the bare vertex in the weak-coupling
limit; (v) it should ensure multiplicative renormalizability of the
DSE in Eq. (\ref{fermDSE_eq});
(vi) the transverse part of the vertex should be specified to 
	ensure gauge-covariance of the DSE.

Ball and Chiu \cite{BC} have given a description of the most general
fermion-photon vertex that satisfies the WTI;
it consists of a longitudinally-constrained (i.e., ``Ball-Chiu'') part
$\Gamma^\mu_{\rm BC}$, which is a minimal solution of the WTI
with no artificial kinematic singularities,
and a basis set of eight transverse vectors $T_i^\mu(k,p)$,
which span the hyperplane specified by
${\cal L}_{\mu\nu}T_i^\nu(k,p) = 0$
(i.e., $q_\nu T_i^\nu(k,p) = 0$), where $q \equiv k-p$.
The minimal longitudinally constrained part of the vertex
will be referred to as the Ball-Chiu vertex and is given by
\begin{equation} \label{minBCvert_eqn}
  \Gamma^\mu_{\rm BC}(k,p) = \frac{1}{2}[A(k^2) +A(p^2)] \gamma^\mu
    + \frac{(k+p)^\mu}{k^2-p^2}
      \left\{ [A(k^2) - A(p^2)] \frac{{\not\!k}+ {\not\!p}}{2}
	      - [B(k^2) - B(p^2)] \right\}\:.
\end{equation}
Note that since neither ${\cal L}_{\mu\nu}\Gamma^\nu_{\rm BC}(k,p)$
nor ${\cal T}_{\mu\nu}\Gamma^\nu_{\rm BC}(k,p)$ vanish identically,
the Ball-Chiu vertex has both longitudinal and transverse components.
The transverse vectors can be conveniently written as \cite{Kiz_et_al}
\begin{eqnarray}
  T_1^\mu(k,p) & = & p^{\mu}(k\cdot q)-k^{\mu}(p\cdot q)\:, 
                                                        \label{T1mu}\\
  T_2^\mu(k,p) & = & \left[p^{\mu}(k\cdot q)-k^{\mu}(p\cdot q)\right]
                     ({\not\! k}+{\not\! p})\:,         \label{T2mu}\\
  T_3^\mu(k,p) & = & q^2\gamma^{\mu}-q^{\mu}{\not \! q}\:,
							\label{T3mu}\\
  T_4^\mu(k,p) & = & q^2\left[\gamma^{\mu}(\not\!p+\not\!k)
			        -p^{\mu}-k^{\mu} \right]
		     - 2i(p-k)^{\mu} k^{\lambda} p^{\nu}
                     \sigma_{\lambda\nu}\:,             \label{T4mu}\\
  T_5^\mu(k,p) & = & -iq_{\nu}{\sigma^{\nu\mu}}\:,      \label{T5mu}\\
  T_6^\mu(k,p) & = & \gamma^{\mu}(p^2-k^2)+(p+k)^{\mu}{\not \! q}\:,
                                                        \label{T6mu}\\
  T_7^\mu(k,p) & = & \frac{1}{2}(p^2-k^2)
		     \left[\gamma^{\mu} ({\not \! p}+{\not \! k})
			     -p^{\mu}-k^{\mu}\right]
                     - i\left(k+p\right)^{\mu}k^{\lambda}p^{\nu}
                       \sigma_{\lambda\nu}\:,           \label{T7mu}\\
  T_8^\mu(k,p) & = & i\gamma^{\mu}k^{\nu}p^{\lambda}
		     {\sigma_{\nu\lambda}}
                     +k^{\mu}{\not \! p}-p^{\mu}{\not \! k}\:, 
                                                        \label{T8mu}
\end{eqnarray}
where we use the conventions $g^{\mu\nu}={\rm diag}(1,-1,-1,-1)$,
$\{\gamma^\mu,\gamma^\nu\}=2g^{\mu\nu}$, and
$\sigma^{\mu\nu}\equiv (i/2)[\gamma^{\mu},\gamma^{\nu}]$.
A general vertex is then written as
\begin{equation} \label{anyfullG_eqn}
  \Gamma^\mu(k,p) = \Gamma_{BC}^\mu(k,p)
    + \sum_{i=1}^{8} \tau_i(k^2,p^2,q^2) T_i^\mu(k,p)\:,
\end{equation}
where the $\tau_i$ are functions that must be chosen to give the
correct $C$, $P$, and $T$ invariance properties.


As previously mentioned, the renormalization procedure is entirely
standard.  One first determines a finite, {\it regularized\/}
self-energy, which depends on both a regularization parameter
and the renormalization point.
One then performs a subtraction at the renormalization point,
in order to define the renormalization parameters $Z_1$, $Z_2$, $Z_3$
which give the full (renormalized) theory in terms of the regularized
calculation.  Consider the
regularized self-energy $\Sigma'(\mu,\Lambda; p)$, leading to the
DSE for the renormalized fermion propagator,
\begin{eqnarray} \label{ren_inv_S}
  {S}^{-1}(p) & = & Z_2(\mu,\Lambda) [\not\!p - m_0(\Lambda)]
    - \Sigma'(\mu,\Lambda; p) \nonumber\\
    & = & \not\!p - m(\mu) - \widetilde{\Sigma}(\mu;p)
      = A(p^2)\not\!p -B(p^2)\:,
\end{eqnarray}
where $\widetilde{\Sigma}(\mu;p)$ denotes the {\it renormalized}
self-energy and the {\it regularized} self-energy is given by
($q\equiv(k-p)$)
\begin{equation} \label{reg_Sigma}
  \Sigma'(\mu,\Lambda; p) = i Z_1(\mu,\Lambda) e^2 \int^{\Lambda}
    \frac{d^4k}{(2\pi)^4} \gamma^\lambda {S}(\mu;k)
      {\Gamma}^\nu(\mu; k,p)
      {D}_{\lambda \nu}(\mu; q)\:.
\end{equation}
Here $D^{\lambda \nu}(\mu; q)$ and $\Gamma^\nu(\mu; k,p)$ denote
the renormalized photon propagator and photon-fermion proper vertex
respectively.
As suggested by the notation
(i.e., the omission of the $\Lambda$-dependence)
renormalized quantities must become independent of the
regularization-parameter as the regularization is removed
(i.e., as $\Lambda\to\infty$) in a renormalizable theory.
The self-energies are decomposed into Dirac and scalar parts,
\begin{equation}
  \Sigma'(\mu,\Lambda; p) = \Sigma'_d(\mu,\Lambda; p^2) \not\!p
		     + \Sigma'_s(\mu,\Lambda; p^2)
  \label{decompose}
\end{equation}
(and similarly for the renormalized quantity,
$\widetilde{\Sigma}(\mu,p)$).
By imposing the renormalization boundary condition,
\begin{equation}
  \left. {S}^{-1}(p) \right|_{p^2 = \mu^2}
  = \not\!p - m(\mu)\:,
\label{ren_point_BC}
\end{equation}
one gets the relations
\begin{equation}\label{ren_BC}
  \widetilde{\Sigma}_{d,s}(\mu; p^2) =
    \Sigma'_{d,s}(\mu,\Lambda; p^2) - \Sigma'_{d,s}(\mu,\Lambda; \mu^2) 
\end{equation}
for the self-energy,
\begin{equation}
  Z_2(\mu,\Lambda) = 1 + \Sigma'_d(\mu,\Lambda; \mu^2)
\label{eq_Z2}
\end{equation}
for the renormalization constant, and
\begin{equation}
  m_0(\Lambda) = \left[ m(\mu) - \Sigma'_s(\mu,\Lambda; \mu^2) \right]
	/ Z_2(\mu,\Lambda)
\label{baremass}
\end{equation}
for the bare mass.
The mass renormalization constant is then defined as
\begin{equation}
  Z_m(\mu,\Lambda) = m_0(\Lambda)/m(\mu)\:,
\label{Z_m}
\end{equation}
i.e., as the ratio of the bare to renormalized mass.
The vertex renormalization, $Z_1(\mu,\Lambda)$, is identical to
$Z_2(\mu,\Lambda)$ as long as the vertex {\it Ansatz\/} satisfies
the Ward Identity; this is how it is recovered for multiplication
into $\Sigma'(\mu,\Lambda;p)$ in Eq. (\ref{reg_Sigma}).


The chiral limit occurs by definition when the bare mass is set to zero
and the regularization is removed, i.e., maintaining $m_0(\Lambda)=0$
while taking the limit $\Lambda\to\infty$.    
Explicit chiral symmetry breaking (ECSB) occurs when the bare
mass $m_0(\Lambda)$ is not zero.
Dynamical mass generation or
dynamical chiral symmetry breaking (DCSB) is said to have occurred
when $M(p^2)\neq 0$ in the absence of ECSB.  As the coupling strength
increases from zero there is a transition to a DCSB phase
at the critical coupling strength $\alpha_c$.
Concisely, the absence of ECSB means that $m_0(\Lambda)=0$ and the
absence of both ECSB and DCSB (i.e., $\alpha<\alpha_c$) means that
$M(p^2)$, $m(\mu)$, and $m_0(\Lambda)$
simultaneously vanish. (Recall that in the notation that we use here,
$M(p^2)\equiv B(p^2)/A(p^2)$ and $m(\mu)\equiv M(\mu^2)$.)
This is the same definition of the chiral limit that is used in
nonperturbative studies of QCD,
see e.g. Refs.~ \cite{TheReview,MiranskReview,FGMS,Rothe}
and references therein.
Obviously, any limiting procedure where we take $m_0(\Lambda)\to 0$
sufficiently rapidly as $\Lambda\to \infty$ will also lead
to the chiral limit.


Let us temporarily indicate explicitly the choice
of renormalization point by a $\mu$-dependence of the renormalized
quantities, i.e., $A(\mu;p^2)\equiv 1/Z(\mu;p^2)$,
$M(\mu;p^2)\equiv B(\mu;p^2)/A(\mu;p^2)$, etc.
Note that Eq.~(\ref{ren_inv_S}) implies that 
\begin{eqnarray}
A(\mu;p^2)&=&Z_2(\mu,\Lambda)-\Sigma'_d(\mu,\Lambda;p^2)
=1-\widetilde\Sigma_d(\mu,\Lambda;p^2) \;, \nonumber\\
B(\mu;p^2)&=&Z_2(\mu,\Lambda)m_0(\Lambda)+\Sigma'_s(\mu,\Lambda;p^2)
= m(\mu)+\widetilde\Sigma_s(\mu,\Lambda;p^2) \;.
\label{AB_at_mu}
\end{eqnarray} 
The renormalization point boundary condition in
Eq.~(\ref{ren_point_BC}) then leads to
$\widetilde\Sigma(\mu,\Lambda;\mu^2)=0$,
or equivalently, to the two boundary conditions
$A(\mu;\mu^2)=1$ and $M(\mu;\mu^2)=B(\mu;\mu^2)=m(\mu)$.  From 
Eq.~(\ref{AB_at_mu}) we have 
\begin{eqnarray}
\left[A(\mu;p^2)/Z_2(\mu,\Lambda)\right]&=&
1-\left[\Sigma'_d(\mu,\Lambda;p^2)/Z_2(\mu,\Lambda)\right] \;,
\nonumber\\
\left[B(\mu;p^2)/Z_2(\mu,\Lambda)\right]&=&m_0(\Lambda)+
    \left[\Sigma'_s(\mu,\Lambda;p^2)/Z_2(\mu,\Lambda)\right]\;.
\label{AB_rescaled}
\end{eqnarray} 
The renormalization group
is the set of renormalization point transformations
which by definition leave the bare quantities of the theory unchanged.
Hence, since $m_0(\Lambda)$ is renormalization point independent
it is clear from Eq.~(\ref{AB_rescaled}) that
  $A(\mu;p^2)/Z_2(\mu,\Lambda)$,
  $B(\mu;p^2)/Z_2(\mu,\Lambda)$,
  $\Sigma'_d(\mu,\Lambda;p^2)/Z_2(\mu,\Lambda)$, and
  $\Sigma'_s(\mu,\Lambda;p^2)/Z_2(\mu,\Lambda)$
are renormalization point independent.
Hence $S(\mu;p)Z_2(\mu,\Lambda)$ is renormalization point
independent.  In shorthand form we can express this as 
$S(\mu;p)\propto 1/Z_2(\mu,\Lambda)$ under a renormalization point
transformation.  Hence, the choice of renormalization point
is equivalent to the choice of scale for the functions $A$ and $B$. 
Similarly, in the general unquenched case \cite{IZ}
under a renormalization point transformation we have in addition 
$D^{\sigma\nu}(\mu;q)
    \propto \xi(\mu)\propto 1/Z_3(\mu,\Lambda)$,
$e(\mu) \propto Z_2(\mu,\Lambda)
    \sqrt{Z_3(\mu,\Lambda)}/Z_1(\mu,\Lambda)$, and
$\Gamma^\nu(\mu;q,p)
    \propto Z_1(\mu,\Lambda)$.
It is straightforward to verify the consistency of these renormalization
point transformations.  For example, from Eq.~(\ref{reg_Sigma})
we see that these scaling properties ensure that $\Sigma' \propto Z_2$
and hence from Eq.~(\ref{ren_inv_S}) that $S(p)\propto 1/Z_2$
as it should.  Thus, since $Z_1=Z_2$, multiplicative renormalizability
will automatically follow if we ensure that $\Gamma\to c \Gamma$ as
$A(p^2)\to cA(p^2)$ and $B(p^2)\to cB(p^2)$.
This behavior is automatic for the Ball-Chiu
part of the proper vertex $\Gamma_{\rm BC}^\nu(\mu;p,k)$
as can be seen from Eq.~(\ref{minBCvert_eqn}).
This consistency is unaffected by considering the
quenched photon propagator case where $Z_3=1$.
Clearly in order to ensure multiplicative renormalizability
for an arbitrary vertex it is necessary to choose the functions
$\tau_i$ to scale in the same way, i.e., $\tau_i\propto c\tau_i$.
This is the precise statement of the restriction that multiplicative
renormalizability imposes on the proper photon-fermion vertex.

{}From the above arguments we see that under a renormalization point
transformation we must have {\em for all} $p^2$
\begin{eqnarray}
  M(\mu';p^2)&=&M(\mu;p^2)\equiv M(p^2) \;, \nonumber\\
  \frac{A(\mu';p^2)}{A(\mu;p^2)}&=&
     \frac{Z_2(\mu',\Lambda)}{Z_2(\mu,\Lambda)}
     =A(\mu';\mu^2)=\frac{1}{A(\mu;\mu'^2)} \;,
\label{ren_pt_transf}
\end{eqnarray}
from which it follows for the fermion propagator that 
$S(\mu';p)/S(\mu;p)
  = Z_2(\mu,\Lambda)/Z_2(\mu',\Lambda)$
in the usual way.
The behavior in Eq.~(\ref{ren_pt_transf}) is explicitly tested
for our numerical solutions.
It is clear from Eq.~(\ref{ren_pt_transf})
that having a solution at one renormalization point ($\mu$)
completely determines the solution
at any other renormalization point ($\mu'$)
without the need for any further calculation.


\subsection*{Example vertex choice}
\label{subsec_vertex}

Our results and conclusions regarding the chiral limit are general and
do not depend on any specific vertex choice, i.e., any specific choice
for the functions $\tau_i$.
We only require that the vertex satisfy the Ward-Takahashi identity
and that it be consistent with multipicative renormalizability.
However, in order to discuss the renormalized finite
solutions it is necessary to present detailed numerical
calculations.  For this purpose we will use as an example the
modified treatment of the Curtis-Pennington vertex introduced in
Ref.~\cite{qed4_hwr}.

Curtis and Pennington published a series of articles
\cite{CPI,CPII,CPIII,CPIV} describing their specification of
a particular transverse vertex term, in an attempt to produce
gauge-covariant and multiplicatively renormalizable solutions
to the DSE.
In the framework of massless QED$_4$, they eliminated
the four transverse vectors which are Dirac-even and must
generate a scalar term.  By requiring that the vertex $\Gamma^\mu(k,p)$
reduce to the leading log result for $k \gg p$ they were led to
eliminate all the transverse basis vectors except $T_6^\mu$, with a
dynamic coefficient chosen to make the DSE multiplicatively
renormalizable.  This coefficient had the form
\begin{equation}
  \tau_6(k^2,p^2,q^2) = -\frac{1}{2}[A(k^2) - A(p^2)] / d(k,p)\:,
\label{CPgamma1}
\end{equation}
where $d(k,p)$ is a symmetric, singularity-free function of $k$ and $p$,
with the limiting behavior $\lim_{k^2 \gg p^2} d(k,p) = k^2$.
[Here, $A(p^2)\equiv 1/Z(p^2)$ is their $1/{\cal F}(p^2)$.]
For purely massless QED, they found a suitable form,
$d(k,p) = (k^2 - p^2)^2/(k^2+p^2)$.  This was generalized to the
case with a dynamical mass $M(p^2)$, to give
\begin{equation}
  d(k,p) = \frac{(k^2 - p^2)^2 + [M^2(k^2) + M^2(p^2)]^2}{k^2+p^2}\:.
  \label{CPgamma2}
\end{equation}
It is clear that this choice scales as described in our discussion
above and is consistent with multiplicative renormalizability.

One refinement in our application of the C-P vertex \cite{qed4_hwr}
in the present work is associated with subtleties in the ultraviolet
regularization scheme.
Although there have been some exploratory studies of
dimensional regularization for the DSE \cite{dim_reg},
this has not yet proven practical in nonperturbative field theory
and momentum cutoffs for now remain the regularization scheme
of choice in such studies.
Naive imposition of a momentum cutoff destroys the gauge covariance
of the DSE because the fermion self-energy integral contains terms,
related to the vertex WTI, which should vanish
but which are nonzero when integrated under cutoff regularization
\cite{dongroberts,BP1}.
Based on these considerations a ``gauge-covariance-improved'' treatment
of the C-P vertex was proposed, which consists of the replacement
(in the quenched approximation) 
  $\Sigma'_d(\mu,\Lambda; p^2) + Z_1(\mu,\Lambda)\alpha \xi/8\pi
  \to \Sigma'_d(\mu,\Lambda; p^2)$.
Full details and the derivation of this modification
can be found in Appendix A of Ref.~\cite{qed4_hwr}.
This quenched approximation correction does
not spoil the scaling needed for multiplicative renormalization
since in the quenched approximation the correction term scales with
$Z_2(\mu,\Lambda)$ as does $\Sigma'$.
We have now fully specified the sample vertex that we use
in our numerical calculations.

For the numerical calculations \cite{TheReview,qed4_hw,qed4_hwr}
the equations are separated into
a Dirac-odd part describing the finite propagator renormalization
$A(p^2)$, and a Dirac-even part for the scalar self-energy, by taking
$\frac{1}{4}{\rm Tr}$ of the DSE multiplied by $\not\!p/p^2$ and 1,
respectively.  The equations are solved in Euclidean space and so
the volume integrals,
$\int d^4k$, can be separated into angle integrals and an integral
$\int dk^2$; the angle integrals are easy to perform analytically,
yielding the two equations which will be solved numerically.

In order to obtain numerical solutions, the final Minkowski-space
integral equations are first rotated to Euclidean space.  They are
then solved by iteration on a logarithmic grid from an initial guess.
The solutions are confirmed to be independent of the initial guess
and are solved with a wide range of cutoffs ($\Lambda$),
renormalization points ($\mu$), couplings ($\alpha$),
covariant gauge choices ($\xi(\mu)$),
and renormalized masses ($m(\mu)$).
As has been reported in detail elsewhere \cite{qed4_hw,qed4_hwr},
choosing the renormalized mass, $m(\mu)$, and then solving
for the bare mass, $m_0(\Lambda)$, leads to extremely well-behaved
finite solutions for $A(p^2)$ and $M(p^2)$, which do not vary
as we take the continuum limit ($\Lambda\to\infty$).
In addition, the renormalization point transformation properties
given in Eq.~(\ref{ren_pt_transf}) have also been explicitly verified
using our numerical solutions, which typically have an accuracy
of better than 1 in $10^4$.

An important outcome of these studies was the observation that all
solutions which were well-behaved in the continuum limit contained
decaying oscillations in the mass function $M(p^2)$ if one looked
sufficiently far into the ultraviolet (in $p^2$).  As a result,
as $\Lambda\to\infty$ for any given solution the values for the bare
mass also presented a decaying oscillatory behaviour \cite{qed4_hwr}. 
Clearly, these oscillating solutions cannot be obtained
from the chiral limit procedure
(i.e., $m_0(\Lambda)=0$ and $\Lambda\to\infty$)
and hence cannot be chirally symmetric solutions.

\section{Results}
\label{sec_results}

\begin{figure}[htb]
  \vskip0.5cm
  \setlength{\epsfxsize}{11.0cm}   
      \epsffile{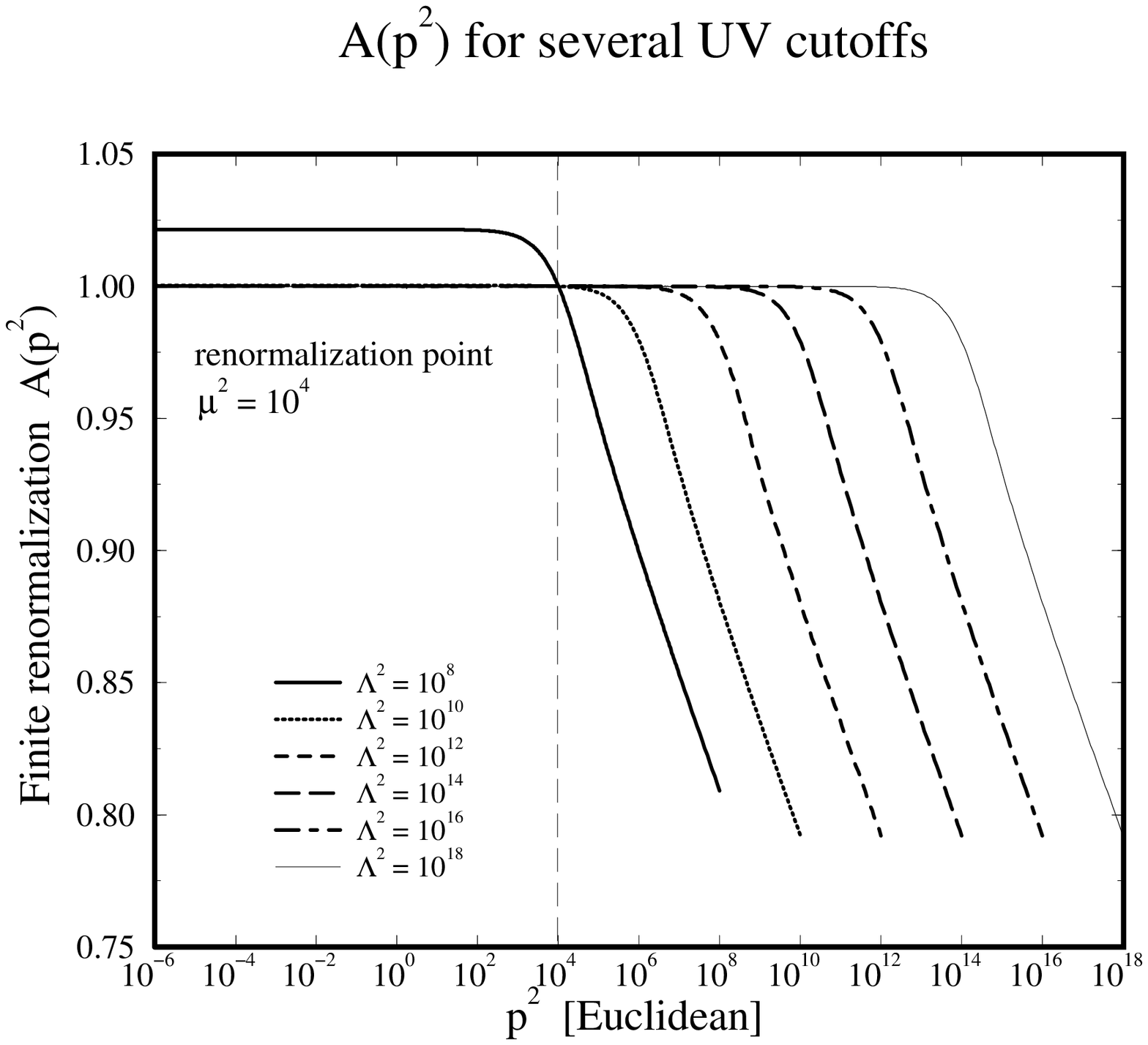}
  \setlength{\epsfxsize}{11.0cm}   
  \vskip0.5cm
      \epsffile{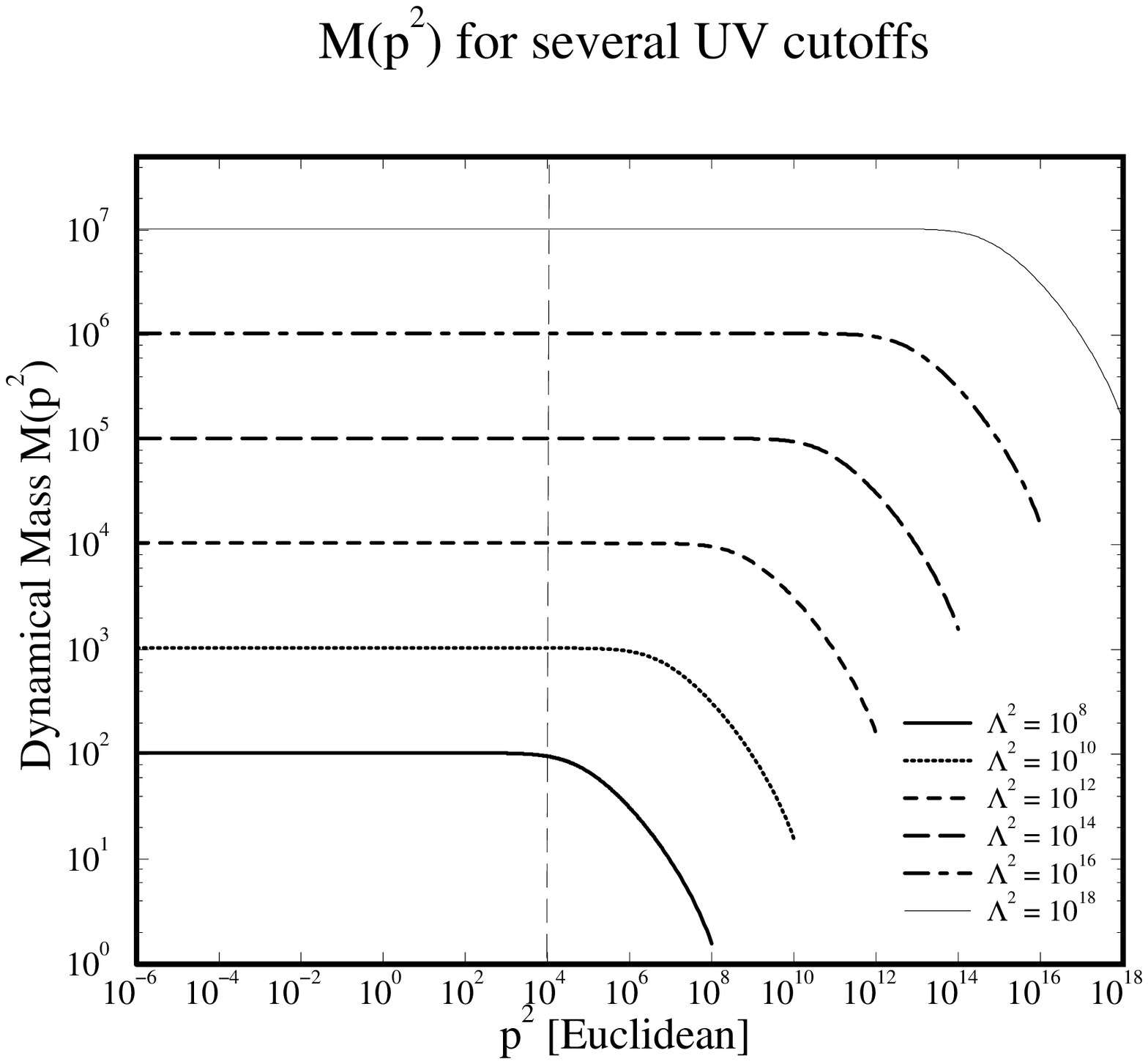}
  \parbox{130mm}{\caption{
        The behaviour of the finite renormalization $A(p^2)$ and the
        mass function $M(p^2)$ as a function of the ultraviolet
        cut-off $\Lambda$ for $m_0(\Lambda)=0$.  These solutions
        were for renormalization point $\mu^2=10^4$,
	coupling $\alpha=1.15$, and gauge parameter $\xi=0.25$.
	Clearly as $\Lambda\to\infty$ we find $A(p^2)\to 1$
	for all $p^2$
	and $M(p^2)$ diverges with $\Lambda$.
  \label{Lambda_limit}}}
  \vspace{0.5cm}
\end{figure}

To motivate our general arguments concerning the chiral limit,
let us consider Fig.~\ref{Lambda_limit},
which shows the non-oscillating solution $A(p^2)$ and $M(p^2)$
for supercritical coupling and with $m_0(\Lambda)=0$,
for a wide range of values of $\Lambda$.
For any given set of parameters
  $\alpha$, $m_0(\Lambda)$, $\mu$, $\xi$, and $\Lambda$
it is always possible to find such a non-oscillating solution.
It is clear from these numerical solutions that in the continuum limit
(i.e., $\Lambda\to\infty$) and for supercritical coupling,
we find $A(p^2)\to 1$ for all $p^2$ 
and a mass function $M(p^2)$ which diverges proportionally
to $\Lambda$.  This divergent behavior of the mass was verified to the
numerical accuracy of our solutions (1 in $10^4$).
Conversely, for subcritical coupling (i.e., $\alpha<\alpha_c$),
the chiral limit does exist and we find simply that
  $A(\mu; p^2) = (p^2/\mu^2)^{-\alpha \xi / 4 \pi}$.
It seems clear from these numerical studies
that above critical coupling there is no finite chiral limit
in the continuum quenched theory in any covariant gauge,
even in the presence of the renormalization procedure.
In the absence of any renormalization program, this divergent behavior
of the mass is well known (see, e.g., Refs.~\cite{CPIII,CPIV}) and is
inevitable since in the absence of ECSB in the quenched theory there
is only one external scale (i.e., $\Lambda$).

While the above conclusions were based on a numerical study with a
specific choice of vertex, it is relatively straightforward to construct
a general argument which applies irrespective of this choice:
  Consider any vertex $\Gamma$ which satisfies
  the WTI and which leads to multiplicative renormalizability
  \cite{qed4_hwr}.  It automatically follows that
  $M(p^2)$ and $A(p^2)/Z_2(\mu,\Lambda)$ are renormalization
  point independent as discussed previously.  We can then define
  dimensionless quantities by appropriately scaling with $\Lambda$,
  i.e., $\widehat\mu\equiv\mu/\Lambda$,
  $\widehat{p}^2\equiv p^2/\Lambda^2$, 
  $\widehat M(\widehat{p}^2)\equiv M(p^2)/\Lambda$,
  $\widehat A(\widehat{p}^2)\equiv A(\mu;p^2)/Z_2(\mu,\Lambda)$,
  and $\widehat{m}_0\equiv m_0(\Lambda)/\Lambda$.
  Note that the renormalization condition $A(\mu;\mu^2)=1$
  automatically determines $Z_2(\mu,\Lambda)$ for a given solution
  for fixed $\Lambda$, (see Eq.~(\ref{eq_Z2})).
  The dimensionless functions $\widehat{M}(\widehat{p}^2)$ and
  $\widehat{A}(\widehat{p}^2)$ of the dimensionless variable
  $\widehat{p}^2$
  can only depend on dimensionless parameters, i.e., $\alpha$, $\xi$,
  $\widehat{\mu}$, and $\widehat{m}_0$. Furthermore, since for any
  fixed $\Lambda$ they are independent of $\mu$ (recall that we are
  working only in the quenched approximation), then it follows
  that they must in turn be independent of $\widehat{\mu}$.
  Now for any finite $\Lambda$ and the choice $m_0(\Lambda)=0$ we
  have $\widehat{m}_0=0$.  Hence solving for any $\mu$ and $\Lambda$
  with $m_0(\Lambda)=0$ allows us to form $\widehat{M}(p^2)$
  and $\widehat{A}(p^2)$, from which we can read off the solutions for
  $A(p^2)$ and $M(p^2)$ for any other $\mu$ and $\Lambda$ with
  vanishing bare mass using the above rules.  
  We see then that the resulting $M(p^2)$ must diverge with $\Lambda$
  as was found numerically.
  To summarize, we see that above critical coupling
  any vertex which satisfies the WTI and leads to multiplicative
  renormalizability will lead to a diverging mass function
  in the continuum limit for quenched QED in four dimensions.  

It is interesting to contrast this result
with the well-known behavior \cite{Politzer} found in QCD,
which leads to a well defined and finite chiral limit in the continuum.
The asymptotic behaviour of QCD has been well-established
in the ultraviolet regime due to asymptotic freedom
in the form of a decreasing running coupling constant $\alpha_s(\mu)$.
This can be used in a renormalization group improved treatment
of the ultraviolet region in the DSE study of DCSB
in the quark propagator \cite{TheReview,MiranskReview,FGMS,WKR}.
       
Given that we know that the chiral limit leads to a divergent mass
function above critical coupling,
what are we then to make of the well-behaved finite solutions?
Our conventional thinking about the chiral limit would imply that
since ECSB should only increase the mass function above that found
in its absence,
then any solution above critical coupling which also has ECSB
in the conventional sense
must also correspond to a divergent mass in the continuum limit.
Thus the finite solutions do not correspond to the chiral limit
nor to any conventional concept of ECSB.

\begin{figure}[htb]
\phantom{nothing here!}
  \vskip1.0cm
  \setlength{\epsfxsize}{11.0cm}
      \epsffile{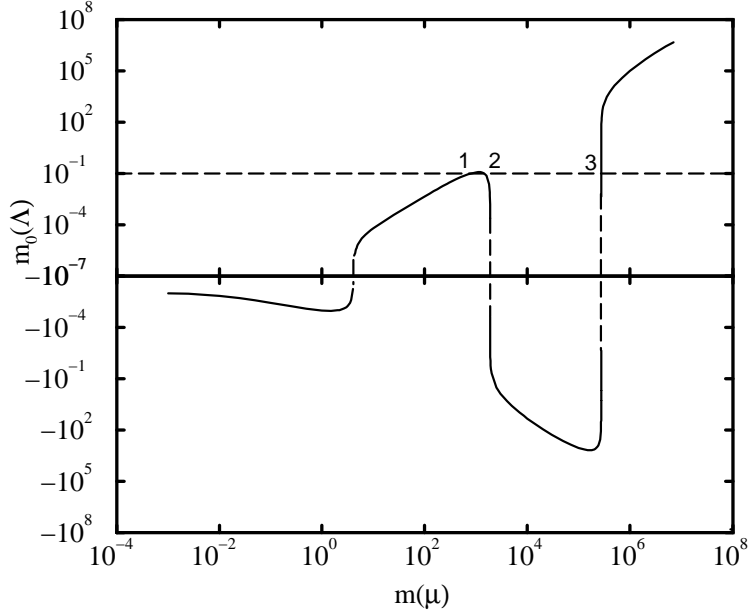}
  \parbox{130mm}{\caption{
        The relationship between the bare mass ($m_0(\Lambda)$)
	and the renormalized mass ($m(\mu)$) for the renormalized
	finite solutions.
        These result from solving for a given $m(\mu)$
	and extracting the corresponding $m_0(\Lambda)$.
	The other parameters for these solutions were
	renormalization point $\mu^2=10^4$, coupling $\alpha=1.25$,
	gauge parameter $\xi=0.25$ and $\Lambda^2=10^{14}$.
        The dashed horizontal line shows, e.g., that
	for $m_0(\Lambda)=0.1$ there are three solutions.  
        (The dashed vertical lines connecting the upper and lower
	curves are merely to guide the eye
	on this back-to-back log scale.)
  \label{mmu_vs_m0}}}
  \vspace{0.5cm}
\end{figure}

\begin{figure}[htb]
  \setlength{\epsfxsize}{11.0cm}
      \epsffile{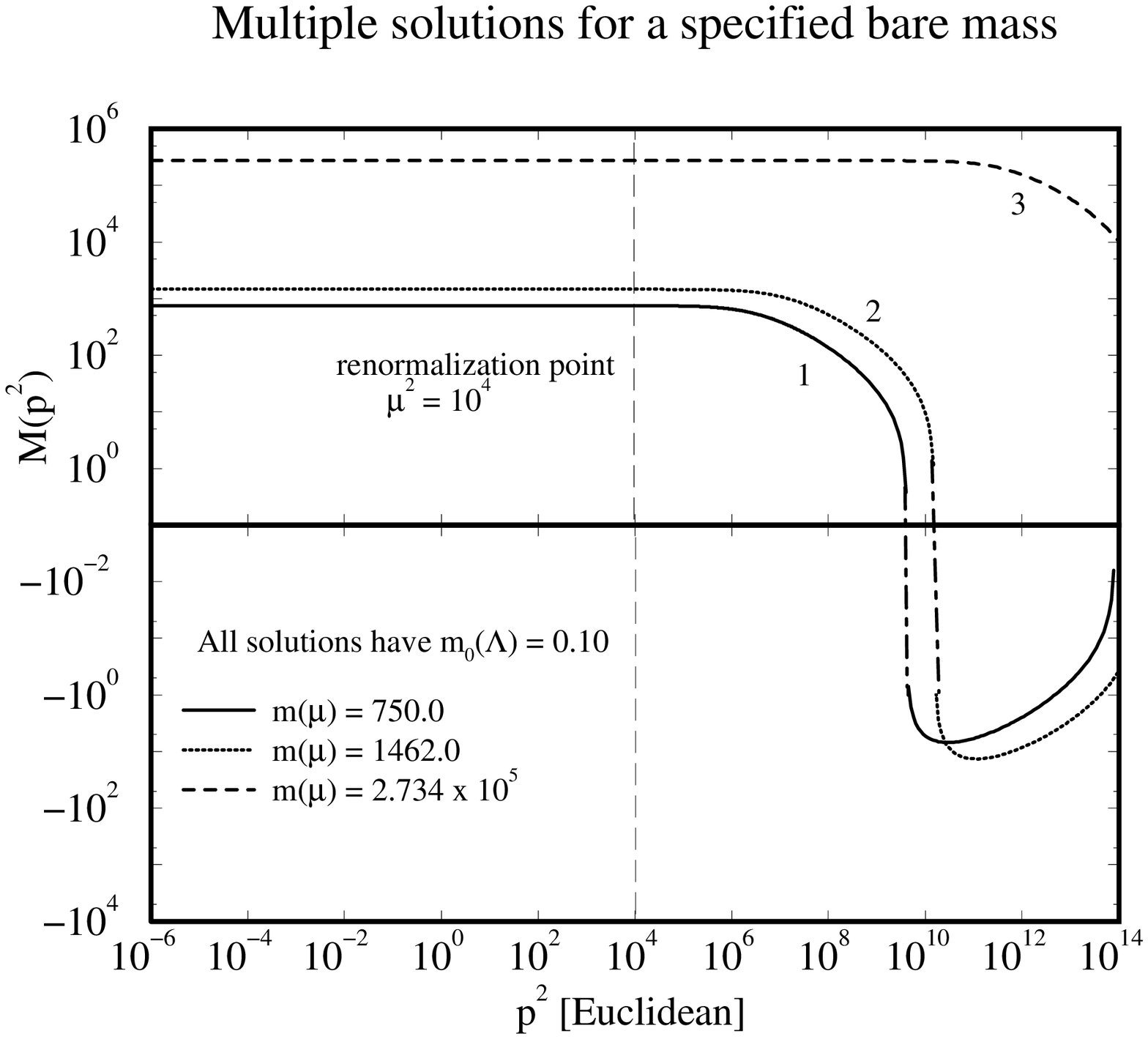}
  \parbox{130mm}{\caption{
        The three renormalized finite solutions
	corresponding to the same bare mass,
	i.e., $m_0(\Lambda)=0.1$, as indicated in Fig.~2.  
        The other parameters for these solutions were
	renormalization point $\mu^2=10^4$, coupling $\alpha=1.25$,
	gauge parameter $\xi=0.25$ and $\Lambda^2=10^{14}$.
        (The short-long dashed vertical lines
	connecting the upper and lower curves
	are merely to guide the eye on this back-to-back log scale.)
  \label{multiple_solutions}}}
  \vspace{0.5cm}
\end{figure}

It has by now probably become clear that a given set of the parameters
  $\alpha$, $m_0(\Lambda)$, $\mu$, $\xi$, and $\Lambda$
actually admits more than one finite solution.
If the renormalization point $\mu$
is chosen to lie far enough in the infrared
that no oscillations in the mass can occur for $p^2<\mu^2$,
then specifying the renormalized mass $m(\mu)$
rather than the bare mass $m_0(\Lambda)$
leads to a unique solution.
This corresponds to the procedure used in obtaining
the numerical solutions in Refs.~\cite{qed4_hw,qed4_hwr}.
We explicitly show this behavior
in Figs.~\ref{mmu_vs_m0} and \ref{multiple_solutions},
where for a given set of parameters
  $\alpha$, $m_0(\Lambda)$, $\mu$, $\xi$, and $\Lambda$
we see that there are distinct solutions.
As $\Lambda\to\infty$ the number of simultaneous solutions
becomes infinite.
This can be understood in the following way:
The oscillations have a period in terms of ln$(p^2)$
which is independent of $\Lambda$ \cite{qed4_hwr}.
Thus the higher is $\Lambda$ the more oscillations can be squeezed
in between the IR and the UV cut-off.
As we increase $m(\mu)$ with everything else fixed,
we push oscillations along the $p^2$ axis.
Each time that a full oscillation is pushed
past the UV cut-off ($\Lambda^2$) there will be
two solutions which give the same value for $m_0(\Lambda)$. 

It seems reasonable to expect that we can also induce these
multiple solutions in QCD by lowering $m(\mu)$ below that which
corresponds to the chiral limit.
This would correspond to a ``negative'' ECSB
and would then again lead to multiple solutions.
A numerical test of this expectation is currently being pursued by
implementing the renormalization program in a QCD-based study of
the quark DSE \cite{qcd_hsw}.
While the DSE for the quark propagator is not yet well-understood
in the IR, the multiple solutions should manifest themselves
in the UV and will not depend on the detailed ans\"{a}tze used
for the QCD propagators and vertex in the IR.

\section{Summary and Conclusions}
\label{sec_conclusions}

  We have studied renormalized quenched strong-coupling QED
  in four dimensions in arbitrary covariant gauge.  
  We saw that there is no finite chiral limit of the renormalized
  theory on general grounds above the critical coupling.
  In addition, we showed that above critical coupling
  for the gauge-covariance-improved treatment
  of the Curtis-Pennington proper fermion-photon vertex,
  there are an infinite number of renormalized finite solutions 
  corresponding to the same bare mass in the continuum limit.
  All of the finite solutions have oscillations and differ in
  how far out on the momentum scale the first oscillation occurs.  

  It seems likely that this behavior for the finite solutions
  is independent of the detailed vertex choice
  and furthermore
  that the same behavior can be induced in QCD
  by forcing the renormalized mass into an unphysical regime
  (by choosing $m(\mu)\equiv M(\mu^2)$ below the value
  corresponding to the chiral limit).
  It also seems likely that unquenching the theory will not remove
  this rather undesirable behaviour in QED$_4$,
  since the running coupling increases with scale
  rather than decreasing as in QCD.
  These latter conjectures are the subject of current investigation
  \cite{qcd_hsw}.
  This study supports the assertion that in four dimensions
  strong coupling QED does not have a continuum limit
  in the conventional sense.


\begin{acknowledgements}
This work was partially supported by the Australian Research Council,
by the U.S. Department of Energy
through Contract No.\ DE-FG05-86ER40273,
and by the Florida State University Supercomputer Computations Research
Institute which is partially funded by the Department of Energy
through Contract No.\ DE-FC05-85ER250000.
This research was also partly supported by grants of supercomputer time
from the U.S. National Energy Research Supercomputer Center
and the Australian National University Supercomputer Facility.
\end{acknowledgements}


\newpage


\newpage

\newpage
\phantom{nothing here!}
\vskip1.0cm


\begin{references}
\bibitem{TheReview}  C.\ D.~Roberts and A.\ G.~Williams,
    {\it Dyson-Schwinger Equations and their Application to Hadronic
    Physics\/}, in
    {\it Progress in Particle and Nuclear Physics, Vol.~33}
    (Pergamon Press, Oxford, 1994), p.~477.
\bibitem{MiranskReview} V.\ A.\ Miranskii,
    {\it Dynamical Symmetry Breaking in Quantum Field Theories},
    (World Scientific, Singapore, 1993).
\bibitem{FGMS} P.\ I.\ Fomin, V.\ P.\ Gusynin, V.\ A.\ Miransky and
    Yu.\ A.\ Sitenko, Riv.\ Nuovo Cim.\ {\bf 6}, 1 (1983).
\bibitem{Rothe} H.\ J.\ Rothe,
    {\it Lattice Gauge Theories: An Introduction},
    (World Scientific, Singapore, 1992).
\bibitem{qed4_hw} F.T.\ Hawes and A.G.\ Williams,
    Phys.\ Rev.\ D {\bf 51}, 3081 (1995).
\bibitem{qed4_hwr} F.~T.\ Hawes, A.~G.\ Williams, and C.~D.\ Roberts,
    hep-ph/9604402,  to be published in Phys.~Rev.~D.
\bibitem{CPI} D.~C.~Curtis and M.~R.~Pennington,
    Phys.\ Rev.\ D{\bf 42}, 4165 (1990).
\bibitem{CPII} D.~C.~Curtis and M.~R.~Pennington,
    Phys.\ Rev.\ D{\bf 44}, 536 (1991).
\bibitem{CPIII} D.~C.~Curtis and M.~R.~Pennington,
    Phys.\ Rev.\ D{\bf 46}, 2663 (1992).
\bibitem{CPIV} D.~C.~Curtis and M.~R.~Pennington,
    Phys.\ Rev.\ D{\bf 48}, 4933 (1993).
\bibitem{Craig_1} C.~D.~Roberts, {\it Schwinger Dyson Equations:
    Dynamical chiral symmetry breaking and Confinement\/}, in
    {\it QCD Vacuum Structure\/}, edited by H.~M.~Fried and
    B.~M\"{u}ller (World Scientific, Singapore, 1993).
\bibitem{BC} J.~S.~Ball and T.~W.~Chiu,
    Phys.\ Rev.\ D{\bf 22}, 2542 (1980); {\it ibid.\/}, 2550 (1980).
\bibitem{Kiz_et_al} A.\ Kizilers\"{u}, M.\ Reenders, and
    M.\ R.\ Pennington, Phys.\ Rev.\ D {\bf 52}, 1242 (1995).
\bibitem{IZ} C.~Itzykson and J.~B.~Zuber, {\it Quantum Field Theory\/},
    (McGraw-Hill, New York, 1980).
\bibitem{dim_reg} L.\ von Smekal, P.~A.\ Amundsen, and R.\ Alkofer,
    Nucl.\ Phys.\ {\bf A529}, 633 (1991);
    M.\ Becker, ``Nichtperturbative Strukturuntersuchungen der
      QED mittels gen\"{a}herter Schwinger-Dyson-Gleichungen
      in Dimensioneller Regularisierung,'' Ph.D.\ dissertation,
      W.~W.~U.\ M\"{u}nster,1995.
\bibitem{dongroberts} Z.~Dong, H.~Munczek, and C.~D.~Roberts,
    Phys.\ Lett.\ {\bf 333B}, 536 (1994).
\bibitem{BP1} A.~Bashir and M.~R.~Pennington,
    Phys.\ Rev.\ D {\bf 50}, 7679 (1994);
    Phys.\ Rev.\ D {\bf 53}, 4694 (1996).
\bibitem{Politzer} H.~D.\ Politzer,
    Nucl.\ Phys.\ {\bf B117}, 397 (1976).
\bibitem{WKR} A.~G.\ Williams, G.\ Krein, and C.~D.\ Roberts,
    Annals of Phys.\ {\bf 210}, 464 (1991).
\bibitem{qcd_hsw} F.~T.\ Hawes, T.\ Sizer, and A.~G.\ Williams,
    in progress.

\end{references}
\end{document}